\input{aipcheck}

\documentclass[
    ,final            
  ]
  {aipproc}

\layoutstyle{6x9}

\begin{document}

\title{Measurements of high-p$_\mathrm{T}$ neutral mesons in $\sqrt{s_\mathrm{NN}}$ = 200~GeV Au+Au and Cu+Cu Collisions at RHIC-PHENIX}
\classification{25.75.-q}
\keywords      {Jet quenching, neutral pion, Relativistic heavy ion collisions}

\author{Tadaaki Isobe for the PHENIX collaboration}{
  address={Graduate School of Science, University of Tokyo, 7-3-1 Hongo,
  Bunkyo-ku, Tokyo 113-0033, Japan}, email={isobe@cns.s.u-tokyo.ac.jp}
}

\begin{abstract}
Measurements from the RHIC experiments show strong suppression 
of high-p$_\mathrm{T}$ hadrons in central Au+Au collisions.
The PHENIX experiment has observed strong suppression of $\pi^0$
and charged hadron yields in central Au+Au collisions for 
p$_\mathrm{T} >$ 5~GeV/$c$ regardless of p$_\mathrm{T}$ and particle
species.  
The observed suppression is interpreted in terms of energy loss of
quarks in a high-density medium. 

The PHENIX Year-4 Au+Au measurements have extended the p$_\mathrm{T}$
reach of the $\pi^0$ spectra to 20~GeV/$c$. 
And the PHENIX Year-5 Cu+Cu measurements performed for the study of
particle production in a lighter system and they can provide the
system-size dependence of the suppression. 
The suppression patterns of $\pi^0$/$\eta$ at high p$_\mathrm{T}$ are
presented for each collision system (Au+Au/Cu+Cu).  
\end{abstract}

\maketitle


\section{Introduction}

The PHENIX experiment~\cite{phenix} has been carried out at the  
Relativistic Heavy Ion Collider~(RHIC) at Brookhaven National
Laboratory to find evidence of phase transition from normal
nuclear matter to Quark Gluon Plasma~(QGP). QGP is a new phase of matter
consisting of de-confined quarks and gluons. 
One of the most exciting results to date at RHIC is that the yield of
$\pi^0$ at high transverse momentum~($p_\mathrm{T}$) in central
$\sqrt{s_\mathrm{NN}}$=200~GeV Au+Au collisions is suppressed compared
to the yield in p+p collisions scaled by the number of underlying
nucleon-nucleon collisions~\cite{bib1}.   
As there is no suppression in d+Au collisions~\cite{bib2}, the
phenomenon is interpreted as a consequence of jet-quenching effect, that
is, hard-scattered partons produced in the initial stage suffer large
energy loss while traversing the dense matter.  

In order to understand the characteristics of the energy loss mechanism
more, the study of the system size dependence and the comparison between
different particle species are important.

\section{Neutral meson measurement at PHENIX}

The PHENIX electromagnetic calorimeter~(EMCal~\cite{emc}) is used to
measure $\gamma$ energy deposit and reconstruct the invariant masses of
$\pi^0$ and $\eta$ via 2$\gamma$ decay mode.

\subsection{Nuclear modification factor}
The amount of suppression can be quantified by nuclear modification
factor~(R$_\mathrm{AA}$). 
R$_\mathrm{AA}$ is the ratio of the measured yield to the expected yield
from p+p result, and is defined as follows:  
\begin{eqnarray}
 R_\mathrm{AA}(p_\mathrm{T}) =
  \frac{d^2N_\mathrm{AA}/dp_\mathrm{T}d\eta}{T_\mathrm{AA}(b)d^2\sigma_\mathrm{NN}/dp_\mathrm{T}d\eta}, 
\end{eqnarray}
\noindent
where the numerator is invariant $\pi^0$ yield in unit rapidity and
denominator is expected yield in p+p collision binary scaled by the
number of underlying nucleon-nucleon collisions~($T_\mathrm{AA}(b)$).
$T_\mathrm{AA}(b)$ is defined as $T_\mathrm{AA}(b) =
N_\mathrm{coll}(b)/\sigma_\mathrm{NN}$, 
where $N_\mathrm{coll}(b)$ is the average number of binary
nucleon-nucleon collisions determined by the distance~$b$ between the center
of two nuclei with the inelastic cross section~$\sigma_{NN}$. 

\subsection{Nuclear modification factor of $\pi^0$ and $\eta$ up to very high p$_\mathrm{T}$}

\begin{figure}[htb]
  \includegraphics[height=.31\textheight]{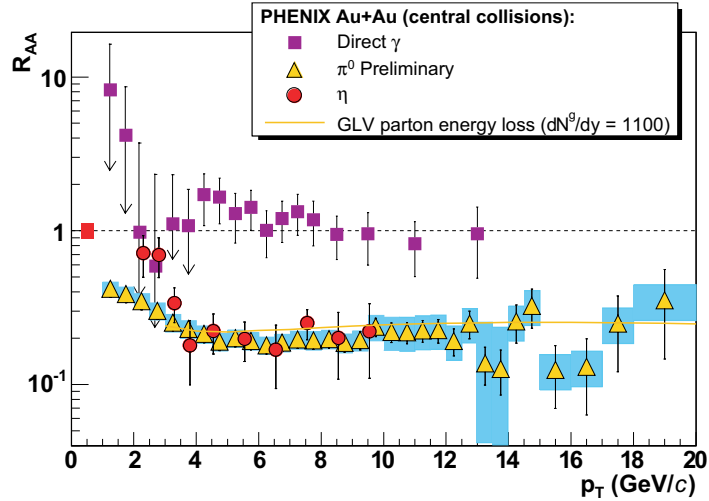}
  \caption{Nuclear modification factor, R$_\mathrm{AA}$ of $\pi^0$
 (triangles), $\eta$ (circles), and direct photon (squares). In addition
 to the statistical and p$_\mathrm{T}$-uncorrelated errors, point-to-point varying
 systematic errors are shown on the data points as boxes. An overall
 systematic error of $T_\mathrm{AA}$ normalization is shown at 1.}
 \label{fig:raa}
\end{figure}

Figure~\ref{fig:raa} shows the preliminary data of $\pi^0$
R$_\mathrm{AA}$ as a function of p$_\mathrm{T}$ with finalized
Year-2 $\eta$, direct photon R$_\mathrm{AA}$ in Au+Au collisions at
$\sqrt{s_\mathrm{NN}}$ = 200~GeV, and theoretical prediction
which employs the GLV model~\cite{200glv}.    
PHENIX Year-3 $\pi^0$ data is used as p+p reference for binary scaling. 
As a result, it is observed that strong $\pi^0$ suppression by a factor
of $\sim$ 5 and it stays almost constant up to 20~GeV/$c$. 
This indicates even a 20~GeV/$c$ pion losses their energy. 
Also the suppression pattern of $\eta$ is found to be similar to that
of $\pi^0$, and this fact supports that the suppression occurs at
partonic level. 
The GLV model describes the strong suppression well and it indicates 
existence of bulk matter where initial gluon density~(dN$^g$/dy) is more
than 1100 in Au+Au collisions at $\sqrt{s_\mathrm{NN}}$ = 200~GeV.

\subsection{Comparison of R$_\mathrm{AA}$ in Au+Au and Cu+Cu collisions}

\begin{figure}[htb]
  \begin{minipage}{0.45\linewidth}
   \begin{center}
    \includegraphics[height=.3\textheight]{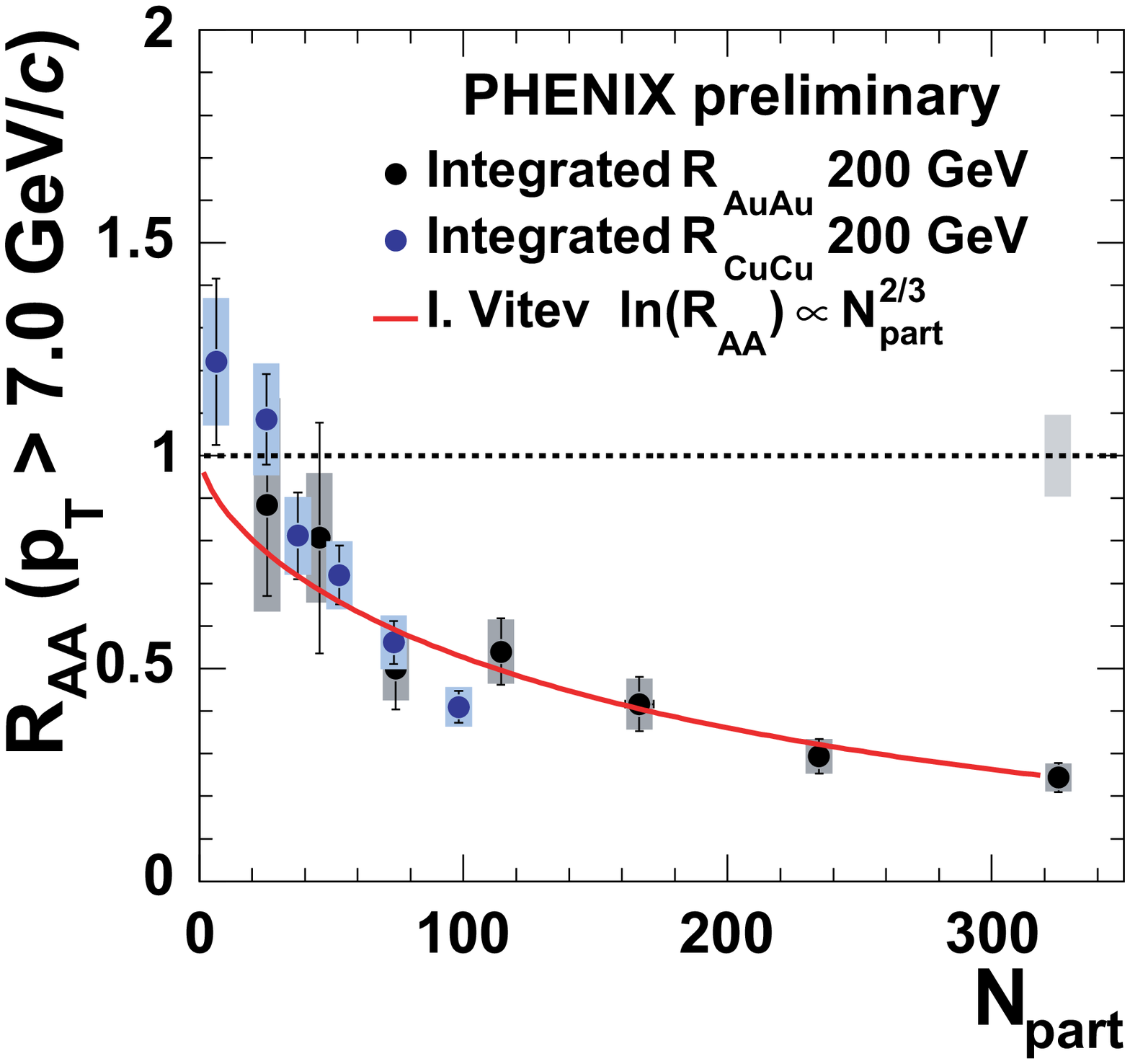}
   \end{center}
  \end{minipage}
  \begin{minipage}{0.45\linewidth}
   \begin{center}
    \includegraphics[height=.25\textheight]{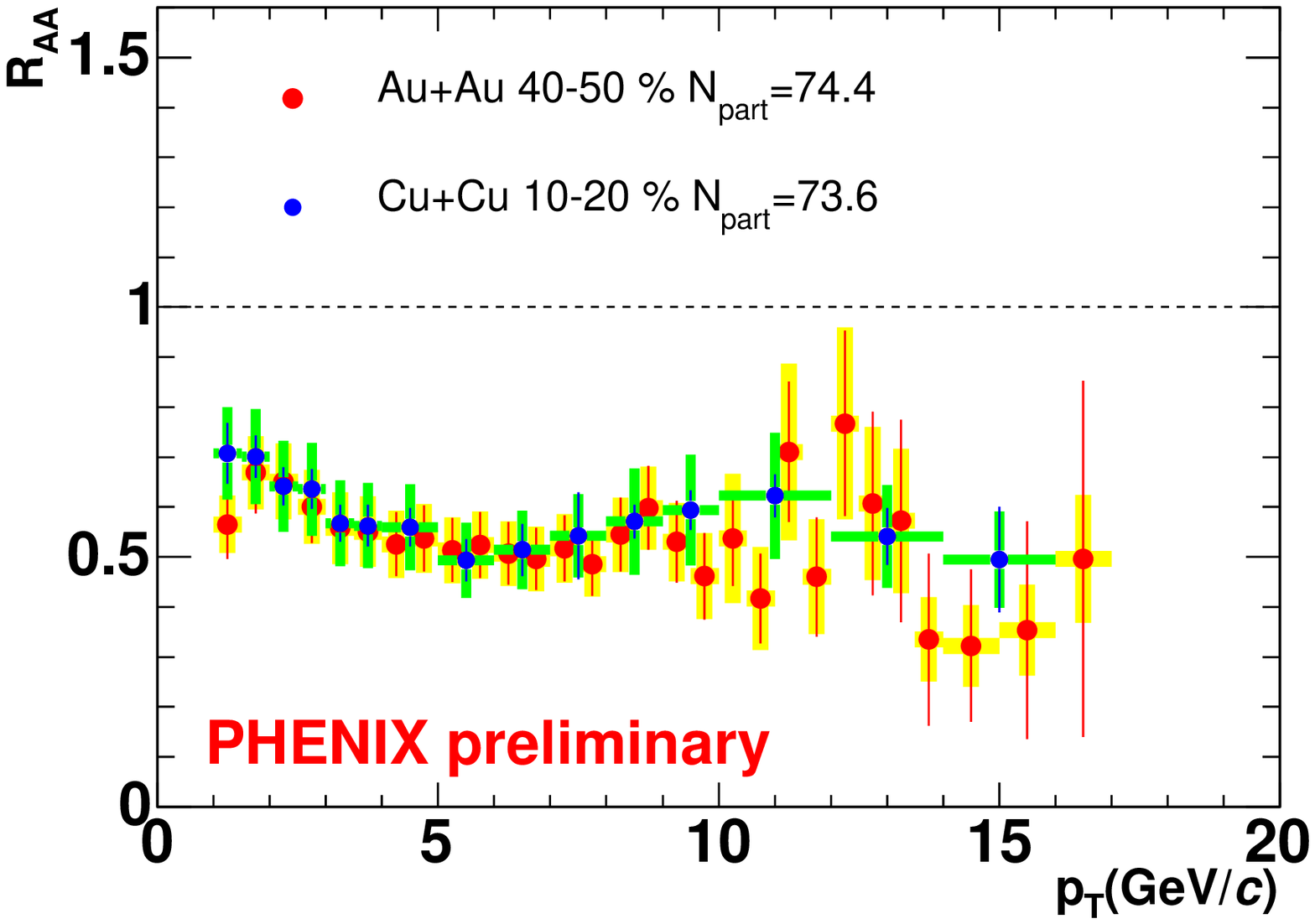}
   \end{center}
  \end{minipage}
 \caption{Left: Integrated R$_\mathrm{AA}$ as a function of the number of
 participant. Right: Comparison of R$_\mathrm{AA}$ between Au+Au and Cu+Cu at
 the similar number of participant (N$_\mathrm{part} \sim$ 74).}
 \label{fig:comp}
\end{figure}

Figure~\ref{fig:comp} shows the comparison of R$_\mathrm{AA}$ in Au+Au
collisions to that in Cu+Cu collisions.
As shown in right panel of Fig.~\ref{fig:comp}, R$_\mathrm{AA}$ in Au+Au
collisions is very similar to that in Cu+Cu collisions at the similar number
of participant.  
Furthermore, integrated R$_\mathrm{AA}$ as a
function of the number of participant (N$_\mathrm{part}$) in Au+Au
collisions is very similar to that in Cu+Cu collisions as shown in the
left panel of Fig.~\ref{fig:comp}. 
The model where energy loss depends on path-length in pure dense matter
describes data well~\cite{200glv}.  

\section{Summary}

$\pi^0$ and $\eta$ are measured in $\sqrt{s_\mathrm{NN}}$ = 200~GeV
Au+Au and Cu+Cu collisions at high-p$_\mathrm{T}$.  
Strong $\pi^0$ suppression by a factor of $\sim$ 5 is observed and stays  
almost constant up to 20~GeV/$c$ in Au+Au collisions.
The R$_\mathrm{AA}$ comparison in Au+Au and Cu+Cu collisions indicates
that the suppression is almost same for similar N$_\mathrm{part}$. 
The universal N$_\mathrm{part}$ scaling of R$_\mathrm{AA}$ reconstructs
the data well.



\end{document}